# Universal method for realization of strong light-matter coupling in hierarchical microcavity-plasmon-exciton systems

Ankit Bisht[1], Jorge Cuadra[1], Martin Wersäll[1], Adriana Canales[1], Tomasz J. Antosiewicz[1,2] and Timur Shegai[1*]

[1]Department of Physics, Chalmers University of Technology, 412 96, Göteborg, Sweden

[2]Faculty of Physics, University of Warsaw, Pasteura 5, 02-093 Warsaw, Poland

* timurs@chalmers.se

**Polaritons are compositional light-matter quasiparticles that arise as a result of strong coupling between a vacuum field of a resonant optical cavity and electronic excitations in quantum emitters. Reaching such a regime is often hard, as it requires materials possessing high oscillator strengths to interact with the relevant optical mode. Two-dimensional transition metal dichalcogenides (TMDs) have recently emerged as promising candidates for realization of the strong coupling regime at room temperature. However, these materials typically provide coupling strengths in the range of 10-40 meV, which may be insufficient for reaching strong coupling with low quality factor resonators. Here, we demonstrate a universal scheme that allows a straightforward realization of strong and ultra-strong coupling regime with 2D materials and beyond. By intermixing plasmonic excitations in nanoparticle arrays with excitons in a $WS_2$ monolayer inside a resonant metallic microcavity, we fabricate a hierarchical system with the combined Rabi splitting exceeding ~500 meV at room temperature. Photoluminescence measurements of the coupled systems show dominant emission from the lower polariton branch, indicating the participation of excitons in the coupling process. Strong coupling has been recently suggested to affect numerous optical- and material-related properties including chemical reactivity, exciton transport and optical nonlinearities. With the universal scheme presented here, strong coupling across a wide spectral range is within easy reach and therefore exploring these exciting phenomena can be further pursued in a much broader class of materials.**

Strong coupling between a vacuum field of a cavity and a quantum emitter can be achieved when the rate of energy exchange between the emitter and the cavity becomes faster than the decoherence rate of both the emitter and the cavity. This regime manifests itself in formation of light-matter hybrid states - polaritons[1,2]. In contrast, weak coupling between a cavity and an emitter does not result in formation of polaritons, but instead leads to the modification of the spontaneous decay rate, known as the Purcell effect[3]. Strong coupling has been previously demonstrated extensively in a plethora of cavity-emitter configurations. Traditionally, high quality factor cavities, such as those based on distributed Bragg reflectors (DBR), whispering gallery modes and photonic crystals, are employed for polaritonic applications[1]. Major drawbacks associated with the use of such cavities are fabrication challenges and the need for cryogenic temperatures. An alternative methodology of utilizing plasmonic nanoparticles for strong coupling applications offers an advantage of smaller mode volumes and open cavity configurations, which in turn allows for room temperature operation[2,4].

On the excitonic side of the problem, semiconductor quantum wells were traditionally employed for the purposes of strong coupling[1]. However, small exciton binding energies of conventional semiconductor materials, such as GaAs, prohibit their use at room temperature[5]. To warrant room temperature operation, organic chromophores and J-aggregates are used ubiquitously for strong coupling purposes. However, they suffer from inhomogeneous broadening, disorder, and photostability issues, thereby hindering their use in practical applications[6-8].

Recently, monolayer TMD semiconductors have emerged as promising alternatives to both semiconductor quantum wells and organic chromophores for exploring polariton physics at room temperature. Monolayer TMDs possess a direct band gap transition, large exciton binding energies, absorption exceeding ~15% at the A-exciton resonance and narrow line width even at room temperature[9-11]. These properties make monolayer TMDs extremely promising for studying polariton-related phenomena. Indeed, plasmonic modes in single nanoparticles have been hybridized with excitons in $WS_2$ and $WSe_2$ in various configurations[12-15] with typical Rabi splitting reaching ~80-120 meV. Similarly nanoparticle lattice modes coupled to $MoS_2$ monolayers have resulted in ~100 meV splitting[16,17]. Strong coupling of both DBR-based and metallic microcavities with monolayer TMDs were shown to produce similar Rabi splitting values of ~20-100 meV[18-20].

More generally plasmon-exciton interactions require strong intermixing between optical and electronic excitations. Such mixing often results in regimes which can be classified as weak, or just at the border between weak and strong[21]. However, to explore rich polaritonic physics, an unambiguous observation of strong or even ultra-strong coupling is required. Typically, this is achieved using solid state architectures[1,5,22], organic microcavities[6-8] or lattice resonances[23,24]. However, these approaches suffer from a necessity to saturate the cavity with a large number of excitonic material(s). Alternatively, strong coupling can be reached by utilizing excitons with high transition dipole moments. However, this reduces the application of strong coupling to a limited class of materials possessing sufficiently high oscillator strengths, such as J-aggregates, quantum dots, perovskites, TMDs, etc.[4]

Here, we present an alternative path to circumvent the problem of reaching strong coupling in nanophotonic systems. The central idea of this study is to realize the strong coupling regime by intermixing plasmons, excitons and microcavity photons into a common polaritonic state. The oscillator strength of plasmonic nanoparticles is orders of magnitude higher than that of TMDs and molecular excitons, which allows the latter to "borrow" the oscillator strength from the former in order to share the same coherent polaritonic states. The microcavities used in our study consist of two 40 nm thick Au mirrors separated by a dielectric spacer of variable thickness, which are filled with plasmonic nanoparticle arrays and monolayers of $WS_2$. By incorporating the system components in this way, we substantially increase the cavity-exciton interaction strength. The energy-momentum dispersion for polaritonic states shows massive Rabi splitting exceeding 500 meV – far into the strong coupling regime. These values are enhanced in comparison to any of the two-component systems - cavity and monolayer $WS_2$ or Au nanoparticles and monolayer $WS_2$. Photoluminescence measurements of the coupled systems show dominant emission from the lower polariton branch, thereby suggesting strong intermixing of excitons with plasmons and cavity photons in our nanophotonic structures. These results provide a universal recipe to reach the strong coupling regime of interaction and pave the way towards exploring its new and emerging applications.

***The concept of oscillator strength borrowing.*** Let us consider a resonant microcavity loaded with a large number of absorbing molecules (see Fig. 1). In the many-emitter strong coupling picture, a large number of emitters coherently contribute to the coupling process to produce the collective Rabi splitting that scales as the square root of the number of involved molecules - $\sqrt{N}$ [25]. Because the process is coherent, the molecular contribution can be thought of as a giant harmonic oscillator with the oscillator strength equal to that of $N$ coherently combined molecules. This in turn implies that all molecules in the coupled system can be replaced by a single entity possessing correspondingly higher oscillator strength, without significantly affecting the mode picture. This concept is illustrated schematically in Fig. 1 (see Methods for further details).

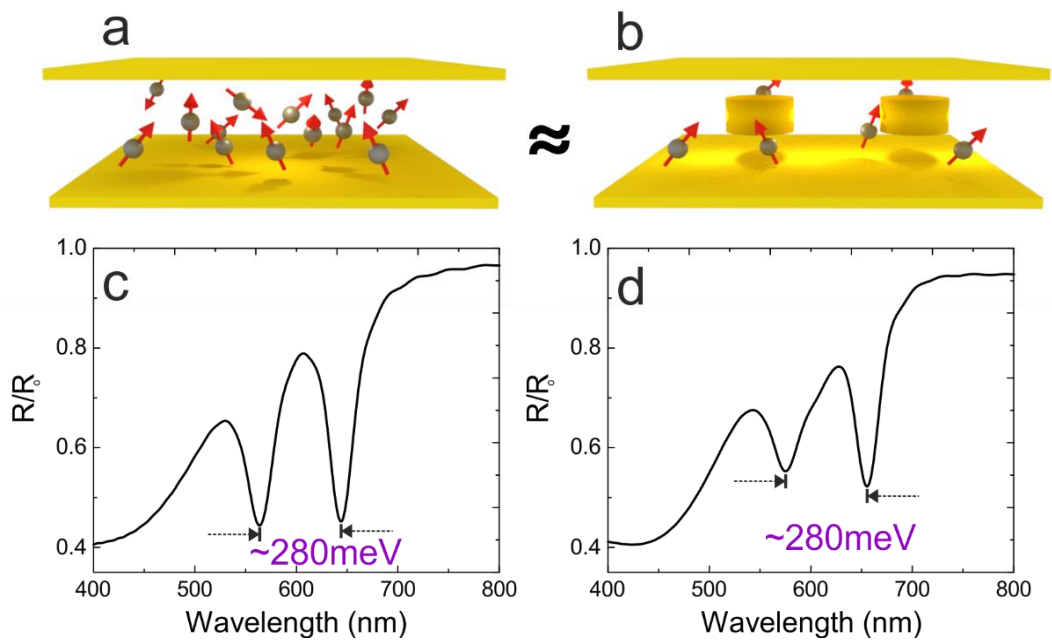

**Figure 1.** Schematic demonstration of electromagnetic similarity between (a) a cavity coupled to a large number of randomly oriented molecular emitters and (b) a cavity coupled to large dipole moment plasmonic nanoparticles with fewer emitters. (c, d) Corresponding calculated reflection spectra show similar Rabi splittings of ~280 meV for both cavity configurations. These observations demonstrate the principle of oscillator strength "borrowing", which allows reaching the strong coupling regime with fewer emitters.

Such an approach is useful for the following reasons. Consider the transition dipole moment for typical excitons. These range in between a few Debyes for fluorophore molecules and few tens of Debyes for quantum dots, perovskites and 2D materials, $\mu_e$=1-50 Debye[4]. Reaching the regime of strong coupling with these excitons requires dense excitonic packing[26] or high-Q microcavity constraints. Contrary to that, plasmonic nanoantennas can easily support transition dipole moments of several thousands of Debyes, owing to the collective electronic nature of plasmonic excitations. In other words, since the absorption cross-section of a typical dye molecule is about ~$10^{-16}$ cm$^2$, while that of a typical plasmonic nanoparticle is ~$10^{-10}$ cm$^2$, a single plasmonic nanoparticle acts as efficiently as approximately ~$10^6$ coherently excited organic molecules in the collective strong coupling picture (provided plasmon and molecule line widths are similar). This is the essence of the oscillator strength "borrowing" - more electron-rich species, such as collective plasmonic resonances in plasmonic nanoparticles, can donate their oscillator strength to electron-poor species, such as single electron excitations in molecular or semiconductor excitons. In this way it is possible to create a microcavity filled up with plasmons and excitons organized in a hierarchical manner resulting in intermixed photon-plasmon-exciton polaritons.

Conceptually our approach is similar to organic-inorganic polaritons pioneered by Agranovich and co-workers[27] and to donor-acceptor polaritons realized experimentally by Barnes group[28] and followed by a number of more recent publications[29-31]. Additionally, microcavities coupled to Au nanospheres were studied previously in the weak-coupling regime, where red-shift of cavity mode was observed due to the presence of the plasmonic particle[32]. Strong coupling of a cavity and a single plasmonic nanorod was recently demonstrated using photoluminescence spectroscopy[33]. Coupling between microcavities and plasmonic lattices at infrared frequencies has also been previously shown[34]. In addition, hybrid systems consisting of plasmonic particles and quantum emitters embedded in microcavities have been studied theoretically[35,36]. Here, we realize similar concepts experimentally by incorporating monolayer WS$_2$ and plasmonic nanoparticle arrays into a common microcavity system.

***Nanoparticle plasmon–microcavity photon polaritons.*** In order to study the coupling within the three-component system, we first realize the coupling between a metallic Fabry-Pérot (FP) microcavity coupled to periodic arrays of Au nanodisks (see Fig. 2a). The thickness of both top and bottom mirrors was set to 40 nm, while the thickness of the cavity was varied from 160 nm to 200 nm. Plasmonic arrays of Au disks (height=20 nm, diameter=80 nm) arranged in square lattices with different particle-to-particle distances ($\Lambda$=200 nm, 250 nm and 300 nm) were fabricated inside the microcavities. The array spacings were chosen such that the dispersive surface lattice modes are in the blue or UV region – far away from the localized plasmon resonances of individual disks.

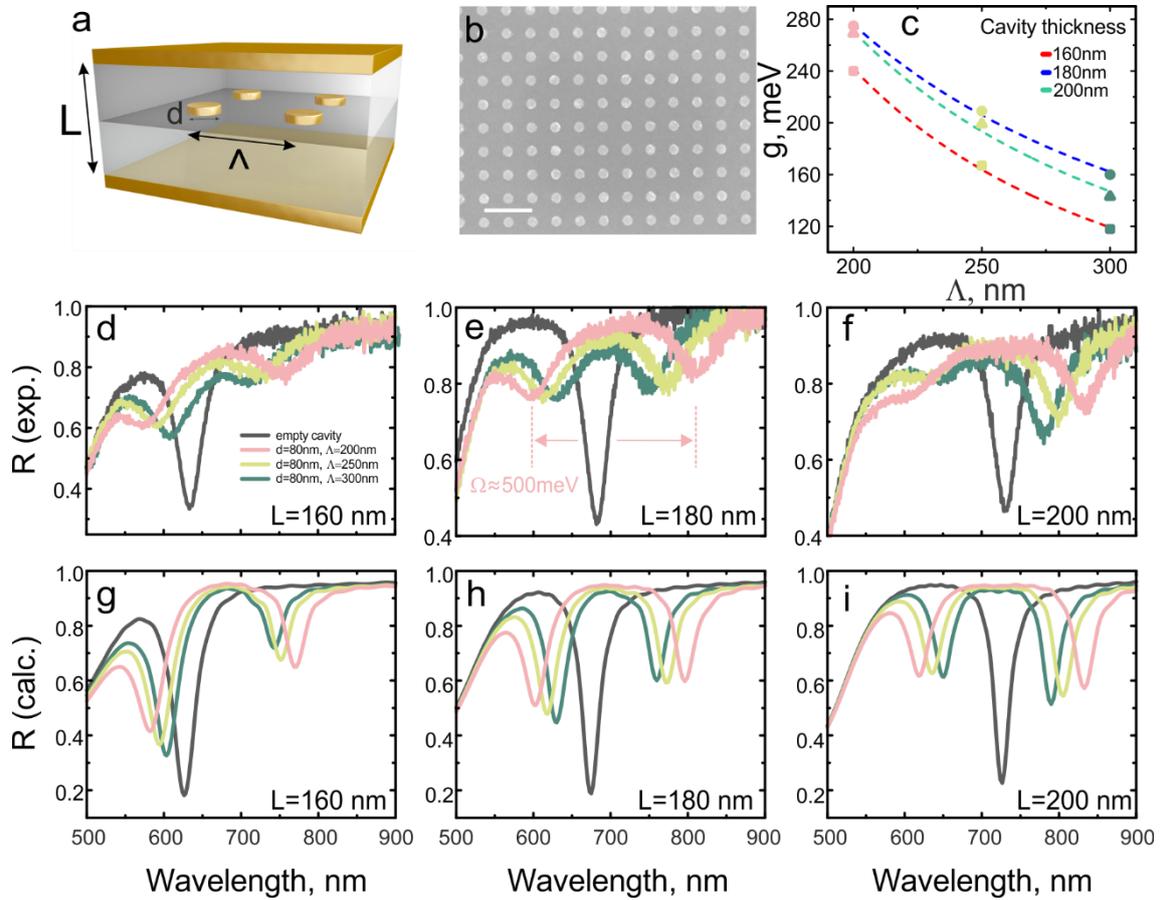

**Figure 2.** Nanoparticle plasmon-photon polaritons in a microcavity system. (a) Schematic of a microcavity coupled to a square lattice of Au nanodisks. Here L represents the cavity thickness, d is the nanoparticle diameter and Λ is the lattice spacing. (b) SEM image of rectangular lattice of Au nanodisks with diameter d=80 nm, height h=20 nm and Λ=200 nm. The scale bar is 400 nm. (c) Coupling strength g as a function of Λ extracted from experimental data using $g = \sqrt{(\omega_+ - \omega_{pl})(\omega_{pl} - \omega_-)}$ where $\omega_+$, $\omega_-$ and $\omega_{pl}$ are energies of upper polariton, lower polariton and nanoparticle localized plasmon. Dashed lines show inverse scaling (g∝Λ$^{-1}$) for all cavity thicknesses. (d-f) Reflection measurements showing microcavity detuning with respect to the nanoparticle plasmon resonance for three different cavity thicknesses: 160 nm (d), 180 nm (e) and 200 nm (f), with each microcavity containing lattices with d=80 nm, h=20 nm and Λ=200 nm, 250 nm and 300 nm. (g-i) FDTD calculations corresponding to experimental reflection in (d-f).

Experimental reflection spectra of bare cavities and cavities loaded with different plasmonic arrays as a function of cavity thickness are shown in Fig. 2d-f. Controlling the thickness of the cavity allows controlling the detuning between the cavity and plasmon resonances. The bare cavity mode splits into two distinct dips in the reflection spectra upon interaction with plasmonic arrays, signaling the formation of polaritonic states. Here, these hybrid states are a mixture of the fundamental cavity mode and localized surface plasmon resonances in Au nanoparticles. These states thus have no excitonic contribution at all.

We further compare experimental findings in Fig. 2d-f with numerical calculations, which were performed using the Finite-Difference Time-Domain (FDTD) method. Exact parameters of the coupled systems extracted from scanning electron microscopy (SEM) images were used in the calculations (example of SEM image is shown in Fig. 2b). The agreement between the calculated (Fig. 2g-i) and experimental (Fig. 2d-f) spectra is remarkable. In this case we find that the plasmon-cavity system is comfortably in the strong coupling regime, since the Rabi splitting significantly exceeds both plasmon and microcavity line widths $\Omega \gg \gamma_{pl}, \gamma_{cav}$ where $\gamma_{pl} \approx$ 200 meV, $\gamma_{cav} \approx$ 150 meV and $\Omega \approx$ 480 meV with the splitting to damping ratio as high as $\Omega/\gamma_{cav} \approx$ 3.2. The plasmon and cavity line width were estimated from the optical data of uncoupled systems. It is important to note that $\Omega \gg \gamma_{pl}, \gamma_{cav}$ is a strict condition for being in the strong coupling regime and satisfying it automatically fulfills weaker conditions, such as $\Omega > (\gamma_{pl} + \gamma_{cav})/2$.

Since the coupling strength is given by $g = \sqrt{N}\mu_{pl}|E_{\text{vac}}|$, where $N$ is controlled by the density of Au nanodisks in the lattices, the transition dipole moment of individual Au disks, $\mu_{pl}$, can be experimentally determined from the measured Rabi splitting values. We observe the $\sqrt{N}$ behavior of mode splitting with the nanoparticle density, similar to the situation of $N$ emitters coherently coupled to a cavity (Fig. 2c). Using the standard expression for the coupling strength, in this case we obtain: $g \approx \mu_{pl}\sqrt{\frac{\hbar\omega}{2\varepsilon\varepsilon_0 L\Lambda^2}}$, where $\Lambda$ is the array spacing and $L$ is the cavity thickness. For experimentally measured values $g \approx$ 280 meV, $\varepsilon$=2.25, $\Lambda$=200 nm and $L$=180 nm, we estimate $\mu_{pl}$ to be about ~1.3x10$^4$ Debyes. This high value is nearly three orders of magnitude higher than a transition dipole moment of A-exciton in monolayer WS$_2$ and any known molecular or quantum dot transition[4]. This is precisely what makes plasmonic nanostructures so useful for reaching the strong coupling regime as we will demonstrate further.

***Plasmon–exciton–microcavity photon polaritons.*** We proceed to explore the composition of the polaritonic mixtures by adding the WS$_2$ monolayer to the system. The plasmon, exciton, and microcavity coupled systems were fabricated by incorporating large area mechanically exfoliated monolayer $WS_2$ flakes inside the microcavities and depositing arrays of Au NPs (see Methods). We intentionally fabricated arrays of plasmonic nanoparticles such that they do not cover the whole area of the flake. This allows to exclude any fabrication-dependent sample-to-sample variations by directly comparing various polariton compositions using the same flake, as shown in the Supplementary Material (SM, Fig. S1). The $WS_2$ monolayers were exfoliated from bulk crystals and transferred inside half microcavities using PDMS stamps (dry transfer technique) [37]. The $WS_2$ monolayer was characterized with optical contrast and Raman spectroscopy (see Fig. S2).

We first compare the two-component (cavity and $WS_2$) versus the three-component (cavity, nanoparticle and $WS_2$) samples using reflection (R) and photoluminescence (PL) spectroscopy (Fig. 3). The experiments were performed for a range of cavity thicknesses (140 nm - 180 nm) to tune the cavity with respect to both the plasmon and exciton resonances. The coupled systems clearly show hybridized reflection dips in both two-component (Fig. 3a) and three-component (Fig. 3d) systems. However, the mode splitting in the latter case is significantly higher, owned to the oscillator strength borrowing from plasmonic nanostructures. FDTD calculations (Fig. 3b,e) performed for these samples agree well with the experimental results. The dielectric function for the monolayer $WS_2$ in this case was extracted from Li et al.[11] We note that for the three-component system, calculations (Fig. 3e) indicate emergence of middle polariton (MP) states, which are nearly absent in experiments (Fig. 3d). We attribute this to differences in polariton line width between experiments and calculations (in experiments it is broader as is evidenced in Fig. 2 and Fig. 3), which may appear as a result of underestimated losses in the Johnson and Christy permittivity of gold[38] that was used in the calculations as well as slight inhomogeneous broadening of nanoparticle plasmons. We also note that control experiments on Au nanodisks deposited on $WS_2$ monolayer outside the cavity do not result in the strong coupling regime as the observed splitting is small in comparison to the nanoparticle plasmon line width (see Fig. S3).

We further investigate the strongly coupled samples using PL spectroscopy. For the case of two-component systems, we find that the PL spectra strongly depend on the cavity thickness (Fig. 3c). This is expected as the cavity thickness determines the cavity-exciton detuning, and thus composition of polaritonic states in this case. For the case of the thinnest cavity (140 nm, black curve in Fig. 3c), PL emission occurs primarily at the A-exciton band of WS$_2$ at around 2 eV, indicating that the polaritonic mixture is mostly of excitonic character (see Hopfield coefficients in Fig. S5a). As the cavity thickness is increased, emission via the lower polariton (LP) branch shifts to lower energy 1.76-1.9 eV depending on the exact cavity thickness. This demonstrates stronger involvement of the cavity mode in the formation of the LP state, consistent with the red-shift of cavity resonance for thicker cavities. Noteworthy, PL emission maxima correspond to the LP maxima measured in reflection (see Fig. 3a-c), in agreement with PL emission from strongly coupled systems[7,39,40].

For the three-component systems, in contrast to two-component counterparts, we find that PL emission is less dependent on the cavity thickness. This can be understood by appreciating the fact that the combined Rabi splittings in the three-component cases are significantly stronger (up to 500 meV) than in two-component systems, thereby implying that detuning between the cavity mode and the plasmon and exciton resonances should play a less important role. Indeed, tuning the cavity thickness from 140 nm to 180 nm does not significantly affect the plasmon-exciton-cavity polariton composition in this case. Here, we again observe that PL signal is dominated by emission from LP states, which occur at 1.6-1.7 eV depending on the cavity thickness. Noteworthy, the LP states in the three-component system lie much below the corresponding two-component counterparts, because of the stronger combined Rabi splitting in this case.

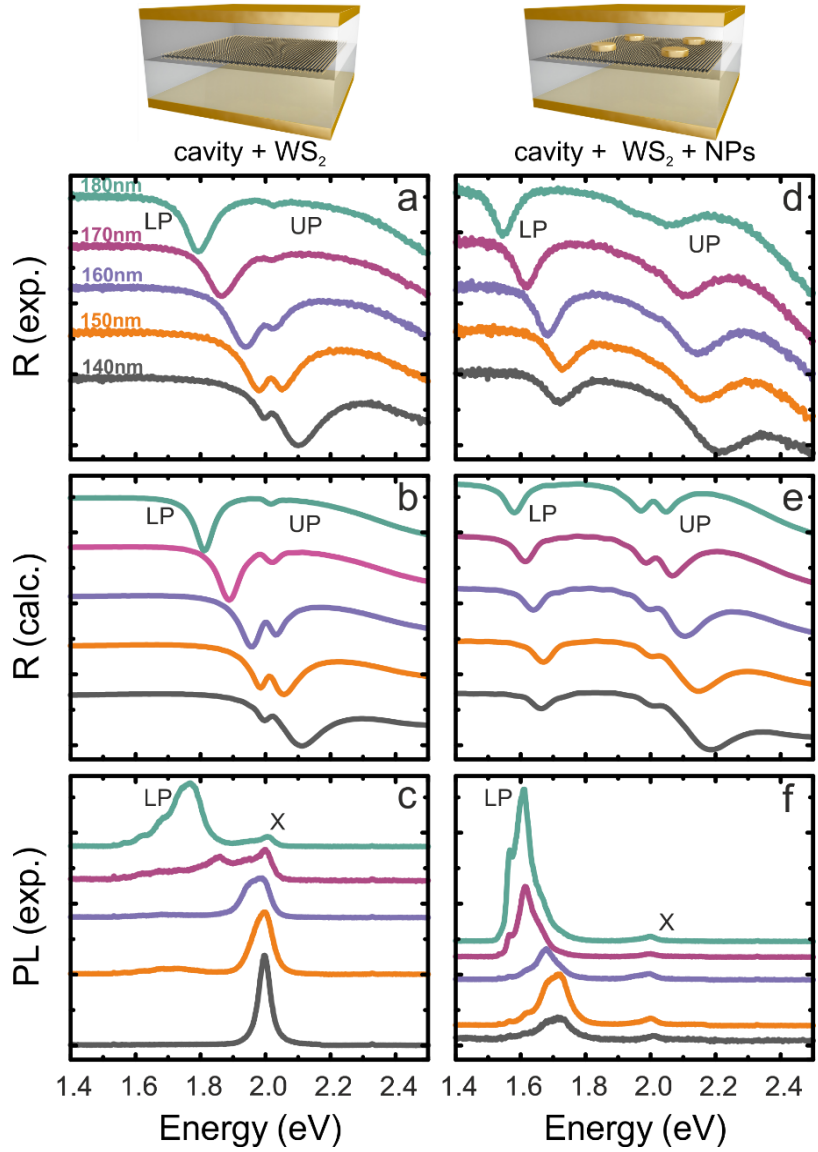

**Figure 3.** Reflectivity (R) and photoluminescence (PL) spectra for samples with different cavity thicknesses (140 nm - 180 nm). (a, d) Experimental near-normal incidence reflection spectra for two-component (a) and three-component (d) systems, correspondingly. (b, e) Calculated reflectivity spectra for two-component (b) and three-component (e) systems, correspondingly. (c, f) Experimental photoluminescence spectra for two-component (c) and three-component (f) systems, correspondingly for excitation wavelength of 532 nm (2.33 eV). Note that PL emission follows the lower polariton branch.

***Polariton dispersion.*** Dispersion of polaritonic modes is an essential characteristic of strongly coupled microcavities. In Fig. 4 we show dispersion of 140, 160 and 180 nm samples as a function of polaritonic composition measured using TE-polarized incidence. Dispersion in reflection was obtained by spectrally scanning the back focal plane of the microscope objective with the liquid crystal filter (LCF, see Methods). The data includes (i) empty cavities, (ii) cavities loaded with monolayer of $WS_2$, (iii) cavities loaded with plasmonic nanoparticles and (iv) cavities loaded with both monolayer WS$_2$ and plasmonic nanoparticles. For the cases when polaritonic mixtures have an excitonic component we complement the reflectivity data with dispersion in PL.

As the cavity thickness is increased, we observe a gradual red shift of the bare cavity mode resonance (Fig. 4a-c), in agreement with the standard microcavity behavior. By introducing the WS$_2$ monolayer into the cavity, the exciton hybridizes with the cavity resulting in UP and LP states, whose exact composition and dispersion depend on the cavity thickness and parameters of the 2D material. PL of these two-component samples show parabolic dispersion and dominant emission via the LP state (Fig. 4d-f), in agreement with previous findings[18,20] and normal incidence data in Figs. 3a-c. Dispersion in the microcavity - WS$_2$ monolayer system shows mode anti-crossing at ~30° (for 160 nm cavity thickness, see Fig. 4e and SI Fig. S5a). The cavity-exciton hybridization results in Rabi splitting of ~75 meV. The obtained splitting is at the border between weak and strong coupling regime, since the cavity and exciton line width are $\gamma_{FP} \approx$ 150 meV and $\gamma_X \approx$ 30 meV correspondingly[20], resulting in $\Omega \lesssim (\gamma_X + \gamma_{FP})/2$=90 meV.

Cavities loaded with plasmonic arrays end up deep in the strong coupling regime, in contrast to two-component cavity-WS$_2$ samples. Dispersion curves in this case show anti-crossing with a substantial Rabi splitting exceeding ~400 meV (Fig. 4g-i). For the 180 nm sample the LP is observed at ~1.7 eV, while the UP is observed at ~2.1 eV although less pronounced. Rabi splittings in this case are slightly smaller than the ones obtained in Fig. 2, because the Au disks diameter in this case was slightly smaller than 80 nm.

When the microcavity, monolayer WS$_2$ and Au nanoparticles are combined together, the LP is observed at ~1.57 eV (at normal incidence), while the UP appears at ~2.15 eV resulting in a massive Rabi splitting of about ~535 meV (see Fig. 4l). Thus, the mode splitting obtained in the three-component system is significantly larger than any of the two-component systems : cavity-WS$_2$ (~75 meV) and cavity-nanoparticles (~400 meV) and even larger than their combination. The latter strongly suggests existence of an additional mechanism of the coupling strength enhancement in this case, which is likely due to the antenna effect as we argue below.

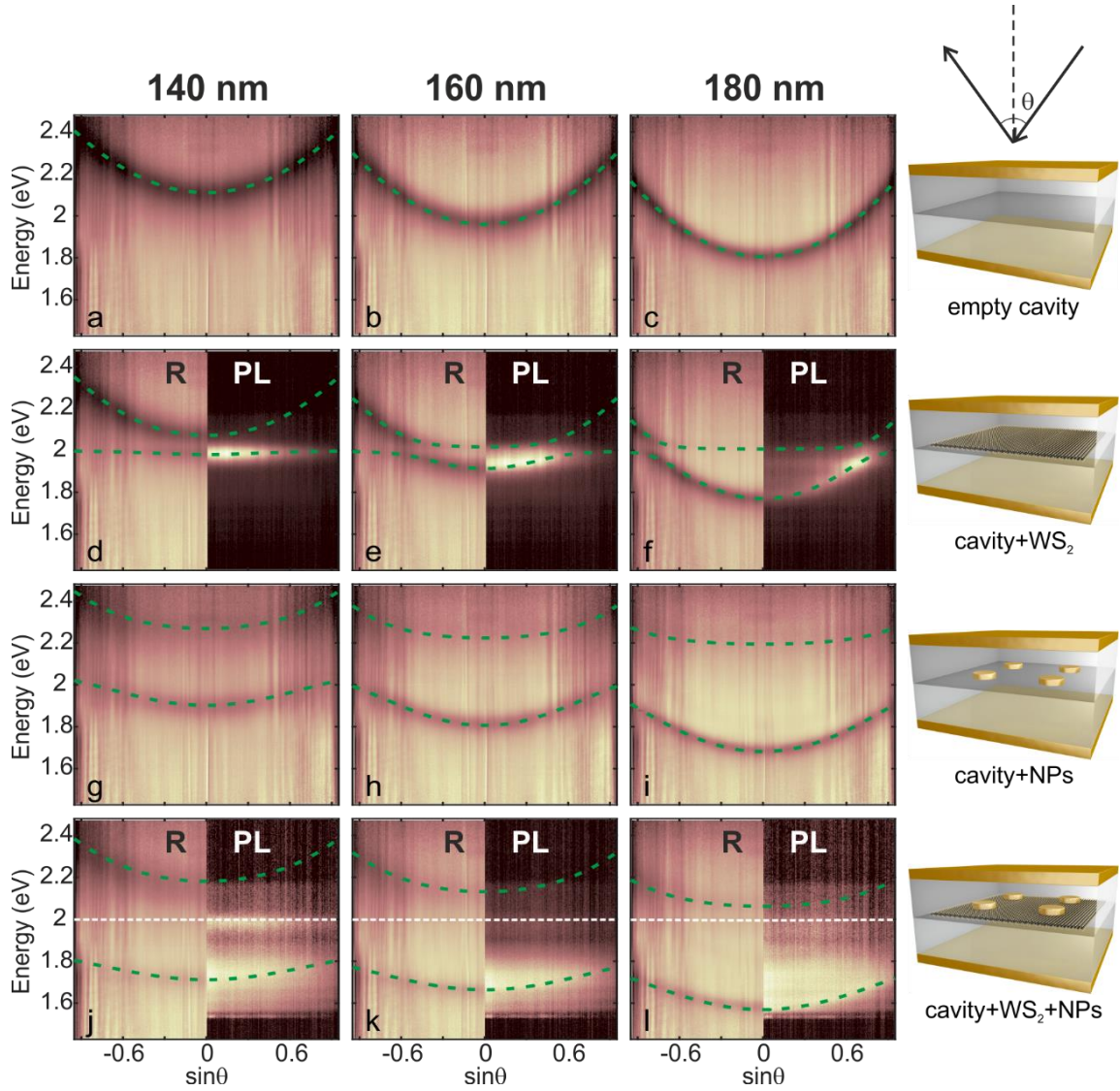

**Figure 4.** Dispersion in reflection and photoluminescence of several microcavity samples as a function of cavity thickness. (a-c) Dispersion for empty cavities, (d-f) Dispersion for cavities loaded with monolayer $WS_2$, (g-i) Dispersion for cavities loaded with plasmonic nanoparticles and (j-l) Dispersion for cavities loaded with monolayer $WS_2$ and plasmonic nanoparticles. Dashed green lines represent UP and LP states extracted from the coupled oscillator model. White horizontal lines in (j-l) represent the position of $WS_2$ exciton.

To improve our understanding of the coupled system, we investigated its PL response. We used a 532 nm (2.33 eV) continuous wave laser to resonantly excite the UP branch. Similarly to reflection measurements, dispersion in PL was obtained by scanning the back focal plane of the microscope objective with the LCF. PL spectra of three samples of different cavity thicknesses are shown in Fig. 4. For 140 nm thickness cavity, the PL signal is heavily dominated by the LP branch (Fig. 4j). We note that the resonance position of the LP branch is *significantly* red-shifted with respect to the resonance of uncoupled $WS_2$ monolayer. Such behavior can be explained by a phonon-assisted population of LP via exciton reservoir in the strong coupling regime[41,42]. The PL data thus points to a significant excitonic contribution to formation of polaritons in the three-component system. This observation is further corroborated by the coupled oscillator model and Hopfield coefficients (see Fig. S5) and by FDTD calculations of absorption spectra of the three-component system (see Fig. S6-S7). As the cavity thickness is increased to 160 nm and 180 nm, the LP emission contribution gets slightly red-shifted as the cavity contribution to LP gradually increases. This is evidenced from dispersion plots in Fig. 4j-l and near-normal incidence spectra in Fig. 3f. For the 180 nm sample, the sample with the most red-detuned cavity mode, PL emission occurs at a slightly higher energy than the LP found in reflection.

To extract additional details we refer to the coupled oscillator model (see Methods and Fig. S5) in Hamiltonian representation. The resulting curves, shown as green dashed lines in Fig. 4, are in good agreement with the experimental results. Such modelling allows extracting the coupling strengths of the corresponding processes (for details see SM). The FP-plasmon coupling strength for the two-component system, $g^{(2)}_{FP-pl}$, turns out to be ~180 meV. This value agrees well with the Rabi splitting obtained for a plasmon-cavity system and the transition dipole moment of a single plasmonic nanoparticle discussed earlier in Fig. 2. On the other hand the FP-WS$_2$ coupling for the two-component cavity-WS$_2$ sample, $g^{(2)}_{FP-X}$, is only about 40 meV, significantly smaller than the corresponding cavity-plasmon coupling.

Importantly, the coupling rates for the three-component system are rather different from the two-component cases. This is because plasmonic particles significantly modify the distribution of electromagnetic energy density in the cavity (see Fig. S4 for further details). As a result of such modification, $g^{(3)}_{FP-X}$ coupling rate reaches as much as 150 meV, which is considerably enhanced in comparison to $g^{(2)}_{FP-X}$. We attribute this to the antenna effect, which modifies the cavity mode such that the electromagnetic energy density is maximized at the position of WS$_2$ monolayer. This observation is in agreement with recent theoretical investigations[36]. At the same time $g^{(3)}_{FP-pl}$ is nearly unaffected in the three-component system and remains to be ~180 meV.

The coupling rate between plasmonic nanoparticles and excitons, $g^{(3)}_{pl-X}$, can be estimated based on near-field FDTD calculations for plasmonic particles inside and outside of the microcavity (see Fig. S4). The analysis shows that $g^{(3)}_{pl-X}$ is relatively small - about 30 meV. We then insert these values into the coupled oscillator model for the three-component system. The obtained curves are in good agreement with the experiment (see dashed green lines in Fig. 4).

Additionally, the coupled oscillator model allows extracting the composition of the polaritonic mixtures (see Fig. S5). In the plasmon-exciton-cavity system we observe a significant contribution of $WS_2$ excitons for all polaritonic branches – UP, middle (MP) and LP. This is further confirmed by FDTD calculations of absorption of the three-component system (Fig. S6) as well as by individual contributions of $WS_2$, Au nanoparticles and cavity mirrors to the total absorption (Fig. S7). In experiments we, however, do not observe strong signatures of MP, which we attribute to underestimated losses in the Johnson and Christy permittivity of gold[38], as mentioned in the discussion of Fig. 3. We note that the overall PL emission behavior, Hopfield coefficients extracted from the coupled oscillator model and FDTD results strongly point to involvement of $WS_2$ excitons into collective mode splitting and formation of macroscopic coherent polaritonic states.

In conclusion, we have shown that by introducing plasmonic nanoparticles inside microcavities, the strong coupling regime with monolayer $WS_2$ system can be readily realized. The main mechanism responsible for these observations is oscillator strength "borrowing". We anticipate that the concept of hierarchical cavity-plasmon-exciton polaritons introduced here may lead to routine observation of strong coupling in microcavities without requiring high exciton densities, albeit at the expense of the reduced exciton character of polaritonic states. Several recent observations suggest that it is collective, not individual Rabi splitting, that is important for a number of physical and chemical processes, such as exciton transport and photochemistry [43-47]. By oscillator strength borrowing one could therefore expect a substantial change in material properties of these strongly coupled emitters, which thereby may lead to new and exciting applications.

**Methods:**

**Fabrication of FP microcavities containing Au nanoparticle arrays and $WS_2$ monolayers:** 170 μm thick glass coverslips (Deckglaser #1) were cleaned in hot acetone, hot isopropanol and Piranha ($H_2O_2$:$H_2SO_4$ 1:3) solution. Bottom microcavity mirror composed of 40 nm Au layer was evaporated using an e-beam evaporator (Kurt J. Lesker PVD225). $SiO_2$ for half-cavities with various thicknesses were then deposited using STS PECVD (Plasma-Enhanced Chemical Vapor Deposition) at 300°C. In order to spatially locate monolayer flakes and align the sample for electron beam lithography Cr/Au (5 nm/25 nm) markers were patterned on the $SiO_2$ using standard UV-lithography process. Large monolayer flakes of $WS_2$ were mechanically exfoliated from bulk crystals (HQ Graphene) on thin PDMS stamps which were later transferred on $SiO_2$ [37]. Monolayer flakes were characterized by PL, optical contrast and Raman scattering (see Fig. S4). Square lattices of Au nanoparticle disks (height 20 nm) with various diameters and pitches were fabricated on the top of $WS_2$ monolayer using standard e-beam lithography. In order to improve the crystallinity of the nanoparticles and remove resist residues, the samples were annealed in an inert atmosphere (Ar/4%$H_2$) for 30 min at 300°C. PMMA layer with same thickness as the bottom $SiO_2$ half cavity was spin coated on top of the Au nanoparticle lattice, followed by baking at 180°C for 5 min. Samples were completed by depositing top microcavity mirror (Au 40 nm) by e-beam evaporation.

**Optical Measurements:** Near-normal reflection spectra were collected using a 20× objective (Nikon, NA=0.45), directed to a fiber-coupled spectrometer and normalized with reflection from a standard dielectric silver mirror. Dispersion relations in reflection were obtained by single shot imaging in the back focal imaging setup. Images were scanned using a liquid crystal filter (LCF) combined with an EM-CCD camera (Andor, iXon). For photoluminescence experiments, the sample was excited by a continuous wave 532 nm laser. PL signal was collected using a 40× objective (Nikon, NA=0.95) and directed to a fiber-coupled 30 cm spectrometer (Andor Shamrock SR-303i) equipped with a CCD detector (Andor iDus 420).

**Coupled Oscillator Model:** To simulate normal modes of the coupled system, we utilize a standard coupled oscillator Hamiltonian in a 3×3 matrix representation. We then solve the eigen value problem. The matrix parameters consists of resonance energies and dissipation rates of all constituent sub-systems (FP cavity - $\omega_{FP}, \gamma_{FP}$, plasmonic nanoparticles - $\omega_{pl}, \gamma_{pl}$, and excitons - $\omega_X, \gamma_X$) together with the different coupling strengths between them ($g^{(3)}_{FP-pl}, g^{(3)}_{FP-X}, g^{(3)}_{pl-X}$). The matrix reads:

$$\widehat{H} = \hbar \begin{pmatrix} \omega_{FP}(\theta) - i\frac{\gamma_{FP}}{2} & g^{(3)}_{FP-pl} & g^{(3)}_{FP-X} \\ g^{(3)}_{FP-pl} & \omega_{pl} - i\frac{\gamma_{pl}}{2} & g^{(3)}_{pl-X} \\ g^{(3)}_{FP-X} & g^{(3)}_{pl-X} & \omega_X - i\frac{\gamma_X}{2} \end{pmatrix} \quad (1)$$

First, we extract the resonance energy and dissipation of a FP cavity as a function of angles. The experimental reflectivity values were fitted using the expression: $\omega_{FP}(\theta) = \omega_{FP}(0) + \alpha \sin^2 \theta$, with $\sin\theta \in [-0.95\ 0.95]$. Thereafter, we model the two-component systems using a 2×2 Hamiltonian matrix representation. Finally the 3×3 Hamiltonian (1) was used to fit experimental data of the three-component system (see Fig. 4j-l). The extracted energies and Hopfield coefficients are shown in Fig. 4 and Fig. S5 correspondingly.

**Finite-Difference Time-Domain (FDTD) Calculations:** Numerical calculations were performed using a commercial FDTD solver: Lumerical, Inc. To obtain the reflection spectra in Fig. 1c, a 150 nm thick molecular layer was assumed to possess a Lorentzian-like dielectric function of the following form:

$$\varepsilon_{total}(f) = \varepsilon_\infty + \frac{f_o \omega_o^2}{\omega_o^2 - 2i\gamma_o \omega - \omega^2}$$

where the Lorentz parameters have a standard meaning and read: $\varepsilon_\infty = 2.1, f_o = 0.05, \omega_o = 2.08\ eV, \gamma_o = 50\ meV$. Such material was then sandwiched between two 40 nm thick Au mirrors, with dielectric function of gold taken from Johnson and Christy [38]. To obtain the reflection spectrum in Fig. 1d, a periodic array of Au disks (d=46 nm, h=20 nm and Λ=200 nm) was placed at the center of the microcavity, while the oscillator strength of the molecular Lorentzian layer was reduced to $f_o = 0.006$.

In Figs. 3b,e the thickness of both mirrors was set to 40 nm, while the thickness of the dielectric (SiO$_2$) layer was varied depending on the spectral requirement of first order cavity resonance. The dielectric function for Au was again taken from Johnson and Christy [38], while SiO$_2$ was modelled as a nearly dispersion free and lossless dielectric with a refractive index of 1.45-1.47 (400 nm to 900 nm). The permittivity of monolayer WS$_2$ was obtained from the literature [11] and used without further modifications. Contributions from all relevant excitonic components were taken into consideration. The WS$_2$ monolayer (thickness 7 Å) was placed at the center of the microcavity. Au nanodisks (d=70 nm and h=20 nm) were placed on top of WS$_2$. Fine meshing was used for monolayer WS$_2$ (0.1 nm), Au disks (1 nm) and the microcavity (1 nm along z-axis) for accurate calculations.

**References:**

1  Khitrova, G., Gibbs, H. M., Kira, M., Koch, S. W. & Scherer, A. Vacuum Rabi splitting in semiconductors. *Nat Phys* **2**, 81-90 (2006).
2  Törmä, P. & Barnes, W. L. Strong coupling between surface plasmon polaritons and emitters: a review. *Reports on Progress in Physics* **78**, 013901 (2015).
3  Purcell, E. M. Spontaneous emission probabilities at radio frequencies. *Physical Review* **69**, 37-38 (1946).
4  Baranov, D. G., Wersäll, M., Cuadra, J., Antosiewicz, T. J. & Shegai, T. Novel nanostructures and materials for strong light-matter interactions. *ACS Photonics*, doi:10.1021/acsphotonics.7b00674 (2017).
5  Kasprzak, J. *et al.* Bose-Einstein condensation of exciton polaritons. *Nature* **443**, 409-414, doi:http://www.nature.com/nature/journal/v443/n7110/suppinfo/nature05131_S1.html (2006).
6  Lidzey, D. G. *et al.* Strong exciton-photon coupling in an organic semiconductor microcavity. *Nature* **395**, 53-55 (1998).
7  Hobson, P. A. *et al.* Strong exciton–photon coupling in a low-Q all-metal mirror microcavity. *Applied Physics Letters* **81**, 3519-3521, doi:doi:http://dx.doi.org/10.1063/1.1517714 (2002).
8  Schwartz, T., Hutchison, J. A., Genet, C. & Ebbesen, T. W. Reversible Switching of Ultrastrong Light-Molecule Coupling. *Physical Review Letters* **106**, 196405, doi:196405 10.1103/PhysRevLett.106.196405 (2011).
9  Mak, K. F., Lee, C., Hone, J., Shan, J. & Heinz, T. F. Atomically Thin ${\mathrm{MoS}}_{2}$: A New Direct-Gap Semiconductor. *Physical Review Letters* **105**, 136805 (2010).
10  Wang, Q. H., Kalantar-Zadeh, K., Kis, A., Coleman, J. N. & Strano, M. S. Electronics and optoelectronics of two-dimensional transition metal dichalcogenides. *Nat Nano* **7**, 699-712 (2012).
11  Li, Y. *et al.* Measurement of the optical dielectric function of monolayer transition-metal dichalcogenides: ${\mathrm{MoS}}_{2}$, $\mathrm{Mo}\mathrm{S}{\mathrm{e}}_{2}$, ${\mathrm{WS}}_{2}$, and $\mathrm{WS}{\mathrm{e}}_{2}$. *Physical Review B* **90**, 205422 (2014).
12  Cuadra, J. *et al.* Observation of Tunable Charged Exciton Polaritons in Hybrid Monolayer WS2–Plasmonic Nanoantenna System. *Nano Letters*, doi:10.1021/acs.nanolett.7b04965 (2018).
13  Wen, J. *et al.* Room-Temperature Strong Light–Matter Interaction with Active Control in Single Plasmonic Nanorod Coupled with Two-Dimensional Atomic Crystals. *Nano Letters* **17**, 4689-4697, doi:10.1021/acs.nanolett.7b01344 (2017).
14  Zheng, D. *et al.* Manipulating Coherent Plasmon–Exciton Interaction in a Single Silver Nanorod on Monolayer WSe2. *Nano Letters* **17**, 3809-3814, doi:10.1021/acs.nanolett.7b01176 (2017).
15  Kleemann, M.-E. *et al.* Strong-coupling of WSe2 in ultra-compact plasmonic nanocavities at room temperature. *Nature Communications* **8**, 1296, doi:10.1038/s41467-017-01398-3 (2017).

16  Liu, W. *et al.* Strong Exciton–Plasmon Coupling in MoS2 Coupled with Plasmonic Lattice. *Nano Letters* **16**, 1262-1269, doi:10.1021/acs.nanolett.5b04588 (2016).
17  Lee, B. *et al.* Electrical Tuning of Exciton–Plasmon Polariton Coupling in Monolayer MoS2 Integrated with Plasmonic Nanoantenna Lattice. *Nano Letters* **17**, 4541-4547, doi:10.1021/acs.nanolett.7b02245 (2017).
18  Liu, X. *et al.* Strong light–matter coupling in two-dimensional atomic crystals. *Nat Photon* **9**, 30-34, doi:10.1038/nphoton.2014.304
http://www.nature.com/nphoton/journal/v9/n1/abs/nphoton.2014.304.html#supplementary-information (2015).
19  Dufferwiel, S. *et al.* Exciton–polaritons in van der Waals heterostructures embedded in tunable microcavities. *Nature Communications* **6**, 8579, doi:10.1038/ncomms9579
http://www.nature.com/articles/ncomms9579#supplementary-information (2015).
20  Wang, S. *et al.* Coherent Coupling of WS2 Monolayers with Metallic Photonic Nanostructures at Room Temperature. *Nano Letters* **16**, 4368-4374, doi:10.1021/acs.nanolett.6b01475 (2016).
21  Antosiewicz, T. J., Apell, S. P. & Shegai, T. Plasmon–Exciton Interactions in a Core–Shell Geometry: From Enhanced Absorption to Strong Coupling. *ACS Photonics* **1**, 454-463, doi:10.1021/ph500032d (2014).
22  Christopoulos, S. *et al.* Room-Temperature Polariton Lasing in Semiconductor Microcavities. *Physical Review Letters* **98**, 126405 (2007).
23  Rodriguez, S. R. K. & Rivas, J. G. Surface lattice resonances strongly coupled to Rhodamine 6G excitons: tuning the plasmon-exciton-polariton mass and composition. *Optics Express* **21**, 27411-27421, doi:10.1364/OE.21.027411 (2013).
24  Väkeväinen, A. I. *et al.* Plasmonic Surface Lattice Resonances at the Strong Coupling Regime. *Nano Letters* **14**, 1721-1727, doi:10.1021/nl4035219 (2014).
25  Tavis, M. & Cummings, F. W. Exact Solution for an $N$-Molecule\char22{}Radiation-Field Hamiltonian. *Physical Review* **170**, 379-384 (1968).
26  Zengin, G. *et al.* Evaluating Conditions for Strong Coupling between Nanoparticle Plasmons and Organic Dyes Using Scattering and Absorption Spectroscopy. *The Journal of Physical Chemistry C* **120**, 20588-20596, doi:10.1021/acs.jpcc.6b00219 (2016).
27  Agranovich, V., Benisty, H. & Weisbuch, C. Organic and inorganic quantum wells in a microcavity: Frenkel-Wannier-Mott excitons hybridization and energy transformation. *Solid State Communications* **102**, 631-636, doi:http://dx.doi.org/10.1016/S0038-1098(96)00433-4 (1997).
28  Andrew, P. & Barnes, W. L. Förster Energy Transfer in an Optical Microcavity. *Science* **290**, 785-788, doi:10.1126/science.290.5492.785 (2000).
29  Coles, D. M. *et al.* Polariton-mediated energy transfer between organic dyes in a strongly coupled optical microcavity. *Nature Materials* **13**, 712, doi:10.1038/nmat3950
https://www.nature.com/articles/nmat3950#supplementary-information (2014).
30  Zhong, X. *et al.* Non-Radiative Energy Transfer Mediated by Hybrid Light-Matter States. *Angewandte Chemie International Edition* **55**, 6202-6206, doi:doi:10.1002/anie.201600428 (2016).
31  Flatten, L. C. *et al.* Electrically tunable organic–inorganic hybrid polaritons with monolayer WS2. *Nature Communications* **8**, 14097, doi:10.1038/ncomms14097
https://www.nature.com/articles/ncomms14097#supplementary-information (2017).
32  Mitra, A., Harutyunyan, H., Palomba, S. & Novotny, L. Tuning the cavity modes of a Fabry–Perot resonator using gold nanoparticles. *Optics Letters* **35**, 953-955, doi:10.1364/OL.35.000953 (2010).
33  Konrad, A., Kern, A. M., Brecht, M. & Meixner, A. J. Strong and Coherent Coupling of a Plasmonic Nanoparticle to a Subwavelength Fabry–Pérot Resonator. *Nano Letters* **15**, 4423-4428, doi:10.1021/acs.nanolett.5b00766 (2015).

34  Ameling, R. & Giessen, H. Cavity Plasmonics: Large Normal Mode Splitting of Electric and Magnetic Particle Plasmons Induced by a Photonic Microcavity. *Nano Letters* **10**, 4394-4398, doi:10.1021/nl1019408 (2010).
35  Peng, P. *et al.* Enhancing Coherent Light-Matter Interactions through Microcavity-Engineered Plasmonic Resonances. *Physical Review Letters* **119**, 233901 (2017).
36  Gurlek, B., Sandoghdar, V. & Martín-Cano, D. Manipulation of Quenching in Nanoantenna–Emitter Systems Enabled by External Detuned Cavities: A Path to Enhance Strong-Coupling. *ACS Photonics* **5**, 456-461, doi:10.1021/acsphotonics.7b00953 (2018).
37  Andres, C.-G. *et al.* Deterministic transfer of two-dimensional materials by all-dry viscoelastic stamping. *2D Materials* **1**, 011002 (2014).
38  Johnson, P. B. & Christy, R. W. Optical-constants of noble-metals. *Physical Review B* **6**, 4370-4379 (1972).
39  Litinskaya, M., Reineker, P. & Agranovich, V. M. Fast polariton relaxation in strongly coupled organic microcavities. *Journal of Luminescence* **110**, 364-372, doi:http://dx.doi.org/10.1016/j.jlumin.2004.08.033 (2004).
40  Wersäll, M., Cuadra, J., Antosiewicz, T. J., Balci, S. & Shegai, T. Observation of Mode Splitting in Photoluminescence of Individual Plasmonic Nanoparticles Strongly Coupled to Molecular Excitons. *Nano Letters* **17**, 551-558, doi:10.1021/acs.nanolett.6b04659 (2017).
41  Savona, V., Andreani, L. C., Schwendimann, P. & Quattropani, A. Quantum well excitons in semiconductor microcavities: Unified treatment of weak and strong coupling regimes. *Solid State Communications* **93**, 733-739, doi:http://dx.doi.org/10.1016/0038-1098(94)00865-5 (1995).
42  Agranovich, V. M., Litinskaia, M. & Lidzey, D. G. Cavity polaritons in microcavities containing disordered organic semiconductors. *Physical Review B* **67**, 085311 (2003).
43  Hutchison, J. A., Schwartz, T., Genet, C., Devaux, E. & Ebbesen, T. W. Modifying Chemical Landscapes by Coupling to Vacuum Fields. *Angewandte Chemie International Edition* **51**, 1592-1596, doi:10.1002/anie.201107033 (2012).
44  Feist, J. & Garcia-Vidal, F. J. Extraordinary Exciton Conductance Induced by Strong Coupling. *Physical Review Letters* **114**, 196402 (2015).
45  Herrera, F. & Spano, F. C. Cavity-Controlled Chemistry in Molecular Ensembles. *Physical Review Letters* **116**, 238301 (2016).
46  Galego, J., Garcia-Vidal, F. J. & Feist, J. Suppressing photochemical reactions with quantized light fields. *Nature Communications* **7**, 13841, doi:10.1038/ncomms13841 (2016).
47  Munkhbat, B., Wersäll, M., Baranov, D. G., Antosiewicz, T. J. & Shegai, T. Suppression of photo-oxidation of organic chromophores by strong coupling to plasmonic nanoantennas. *arXiv preprint* **arXiv:1802.06616** (2018).

# Supplementary Material for

# Universal method for realization of strong light-matter coupling in hierarchical microcavity-plasmon-exciton systems

Ankit Bisht[1], Jorge Cuadra[1], Martin Wersäll[1], Adriana Canales[1], Tomasz J. Antosiewicz[1,2] and Timur Shegai[1*].

[1]Department of Physics, Chalmers University of Technology, 412 96, Göteborg, Sweden

[2]Faculty of Physics, University of Warsaw, Pasteura 5, 02-093 Warsaw, Poland

* timurs@chalmers.se

**Content:**

1. Optical bright-field image of the sample.
2. Raman scattering measurements on $WS_2$ flakes inside FP cavity.
3. Reflection spectrum of Au nanodisks on $WS_2$ monolayer outside the cavity.
4. FDTD calculation of near-field profiles for 160 nm cavity.
5. Coupled oscillator model and Hopfield coefficients for different cavities.
6. Total absorption of the three-component system and absorption in individual components.
7. Spatially-resolved absorption in monolayer $WS_2$ in the three-component system.

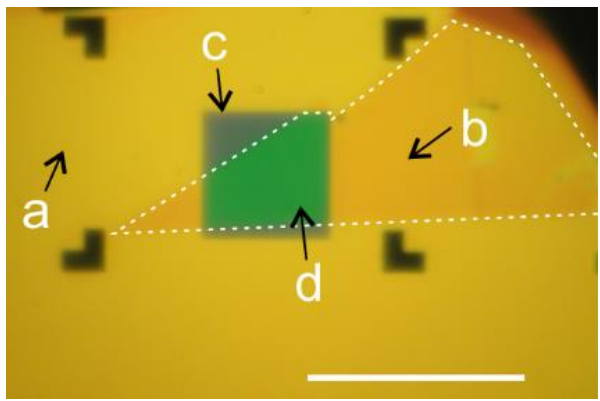

**Figure S1.** Optical bright-field image of the sample. The white dashed lines show the outline of monolayer WS$_2$. Scale bar = 100 μm. Sample configuration allowed measuring all possible cavity compositions, including: empty cavity (region **a**), two-component (cavity + WS$_2$) system (region **b**), two-component (cavity + Au NPs) system (region **c**) and three-component (cavity + Au NPs + WS$_2$) system (region **d**) within a *single sample*. This greatly facilitates comparison between different cavity compositions.

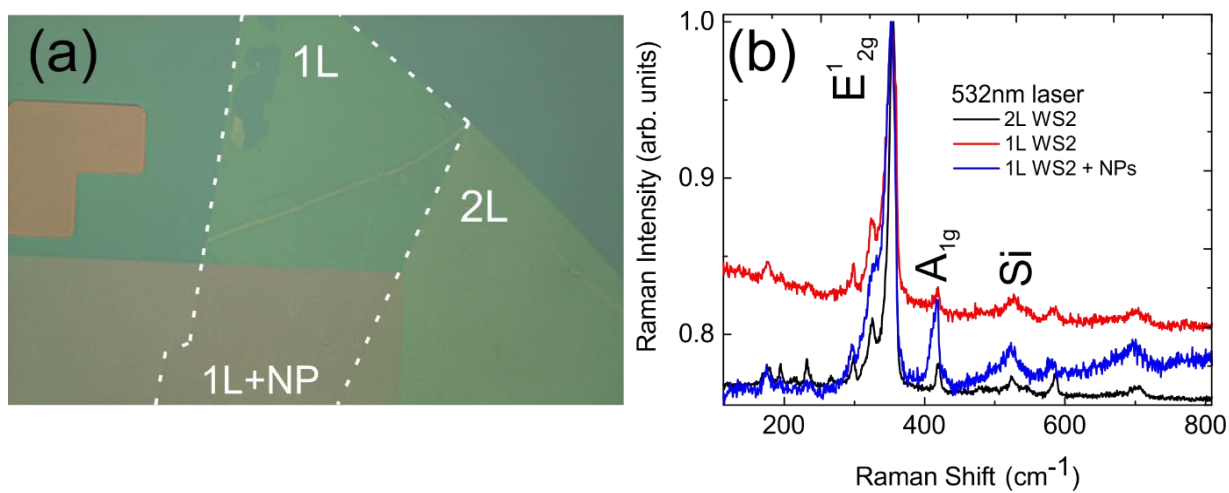

**Figure S2.** (a) Bright field optical image of the fabricated sample consisting of WS$_2$ mono- and bi-layer embedded inside a Fabry-Pérot microcavity together with Au NPs. Monolayer WS$_2$ is marked with white dashed lines. (b) Raman spectroscopy performed on mono-, bi-layer WS$_2$ and monolayer with NPs using a 532 nm excitation laser. The peaks agree well with the Raman spectrum reported in the literature[1] with A$_{1g}$ peak showing a blue shift of ~1 cm$^{-1}$ from mono- to bi-layer. Raman measurements provide additional evidence of WS$_2$ being undamaged by processing steps.

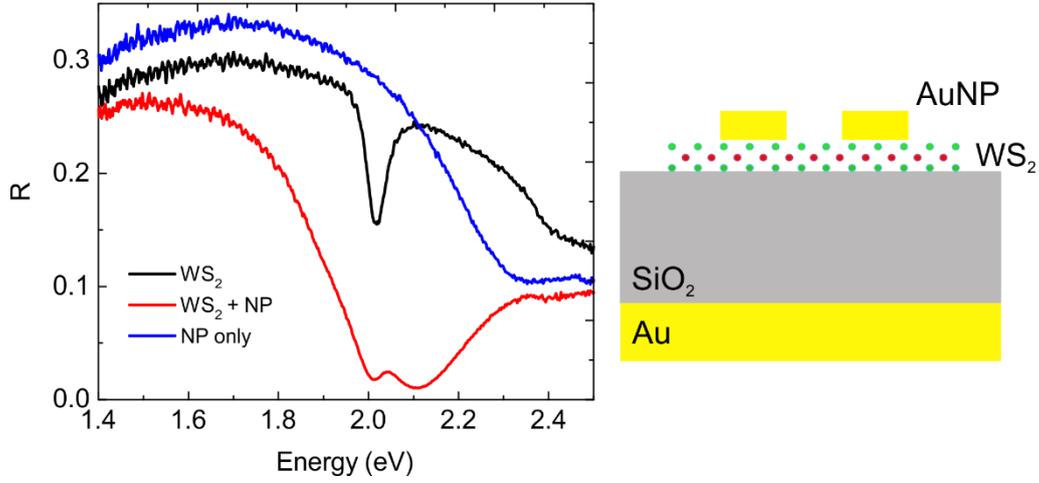

**Figure S3.** Reflection spectra of Au nanodisk arrays coupled to A-exciton in WS$_2$ monolayer without the FP cavity top mirror. SiO$_2$ thickness is 85 nm and measurement was performed using a 50× objective after annealing the sample in Ar/H$_2$ atmosphere for 30 min at 300 $^0$C. The plasmon-exciton coupling in this case can be roughly estimated from: $g = \sqrt{(\omega_+ - \omega_X)(\omega_X - \omega_-)}$. Using measured values for $\omega_+$=2.11 eV, $\omega_-$=2.013 eV and $\omega_X$=2.017 eV, one obtains the coupling strength of nearly $g \approx$20 meV, which is in the weak coupling regime for a typical plasmon line width of 200 meV in this case. Nevertheless, when the cavity is completed with the top mirror, the three-component system ends up deep in the strong coupling regime.

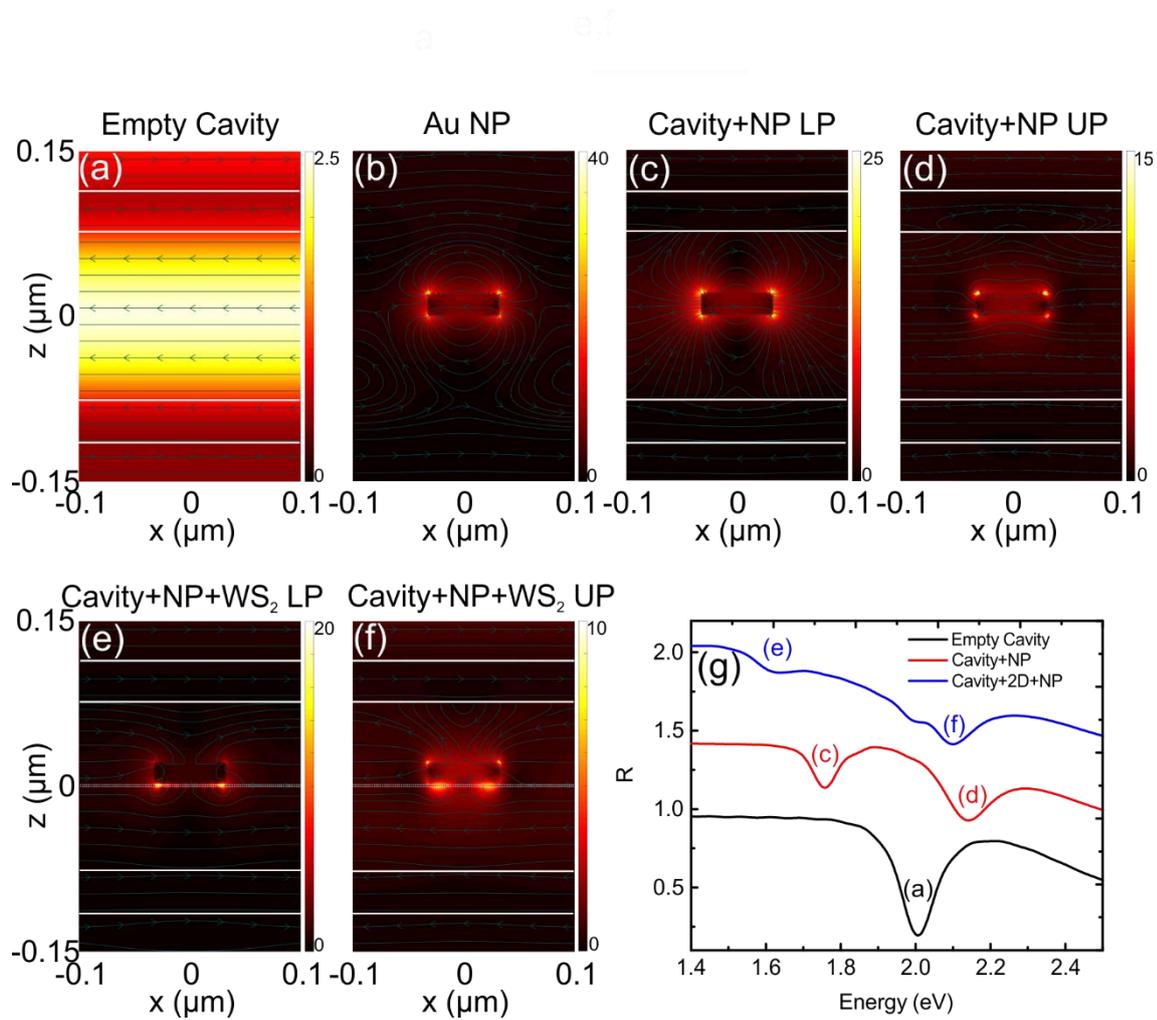

**Figure S4.** (a-f) Near-field profiles for several representative system configurations. (a) Empty cavity with L=160 nm, (b) array of Au NPs outside of the microcavity, (c-d) lower and upper polariton of the cavity and NP system, (e-f) lower and upper polariton of the three-component system and (g) corresponding simulated reflection spectrum with spectral positions where the field profiles were obtained. In panels (a, c-f) the white lines mark the position of Au mirror interfaces, while white dashed lines and black solid lines mark the position of WS$_2$ monolayer and the E-field lines correspondingly.

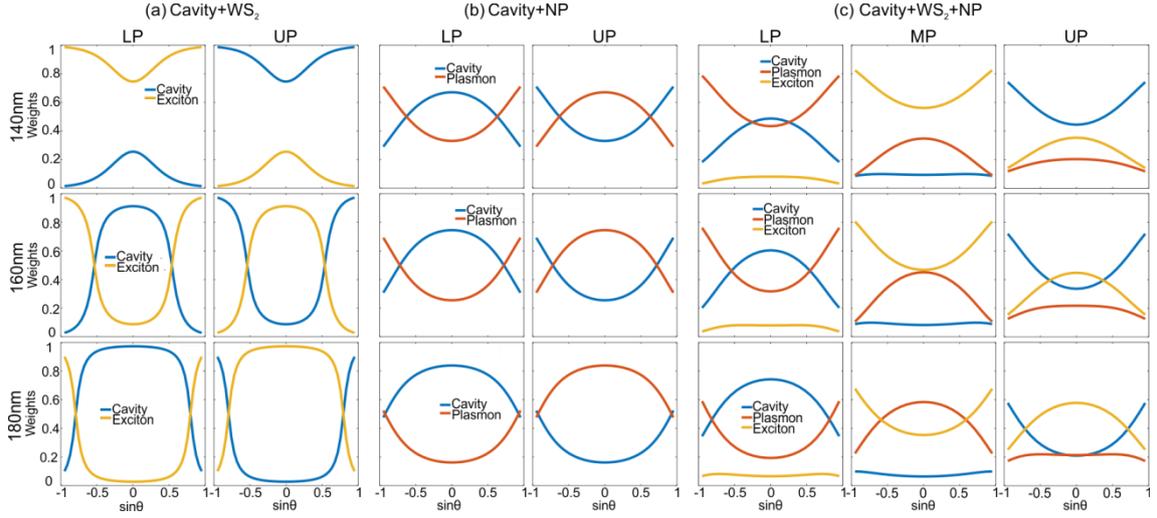

**Figure S5.** Hopfield Coefficients extracted from the coupled oscillator model. The coefficients show contributions of the system components to the composition of various polariton branches in: (a) Cavity + WS$_2$, (b) Cavity + NP and (c) Cavity + WS$_2$ + NP system. In (c) the contribution of excitons to the LP for different cavities is nearly 10% (in agreement with FDTD calculations in Fig. S6 and S7). The parameters used and respective coupling strengths obtained from coupled oscillator model are as follows: (a) Cavity + WS$_2$: $g^{(2)}_{FP-X} = 40\ meV$, $\omega_X = 2.016\ eV$, (b) Cavity + NP: $g^{(2)}_{FP-pl} = 180\ meV$, $\omega_{pl} = 2.1\ eV$, and (c) Cavity + NP + WS$_2$: $g^{(3)}_{FP-pl} = 180\ meV$, $g^{(3)}_{FP-X} = 150\ meV$, $g^{(3)}_{pl-X} = 30\ meV$, $\omega_{pl} = 1.9\ eV$, $\omega_X = 2.016\ eV$. With respective line widths for cavity, exciton and plasmon being: $\gamma_X = 30\ meV$, $\gamma_{FP} = 150\ meV$, $\gamma_{pl} = 200\ meV$. Dispersion of empty cavities was modeled assuming a standard parabolic lineshape: $\omega_{FP}(\theta) = \omega_{FP}(0) + \alpha \sin^2\theta$ with $\alpha$=0.3281, 0.3681 and 0.4081 for cavity with thicknesses of 140 nm, 160 nm and 180 nm, correspondingly.

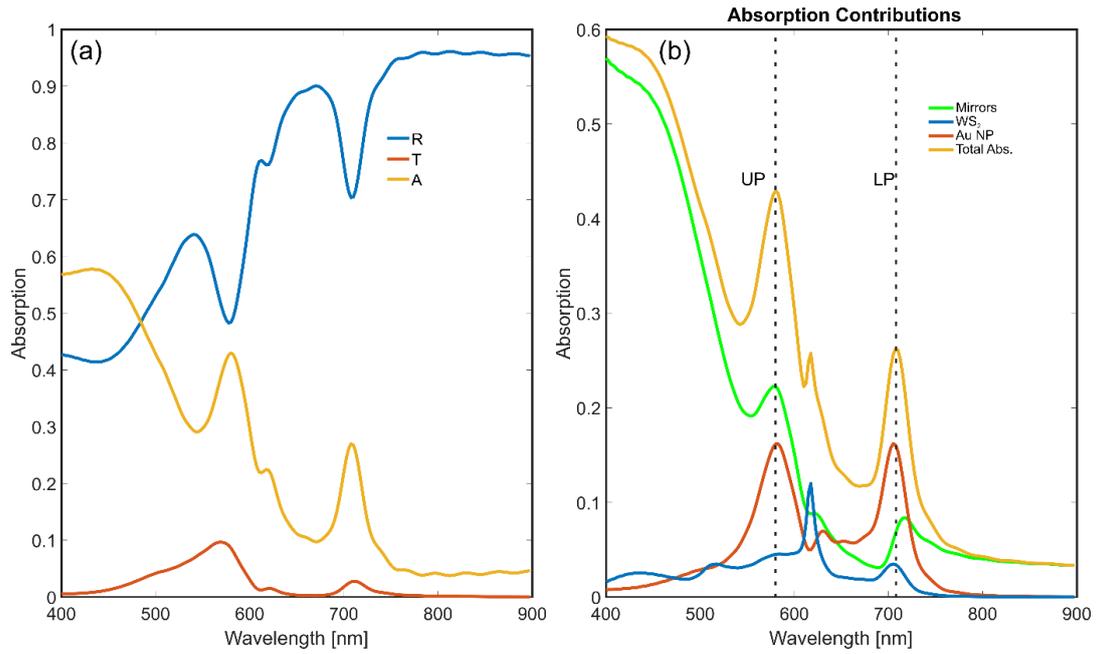

**Figure S6.** (a) Transmission, reflection and absorption spectra for the three-component system (cavity thickness 140 nm) calculated using FDTD. Rabi Splitting is clearly observed in all spectra. (b) Total absorption spectrum for the coupled system from (a) and its decomposition into absorption contributions in mirrors, $WS_2$ and Au NPs. Positions of UP and LP are marked with vertical dashed lines. Splitting in absorption is present in all the individual components of the coupled system. In addition, the absorption of $WS_2$ monolayer shows a peak at the uncoupled $WS_2$ exciton, pointing to a significant excitonic contribution to MP, in agreement with Hopfield coefficient analysis.

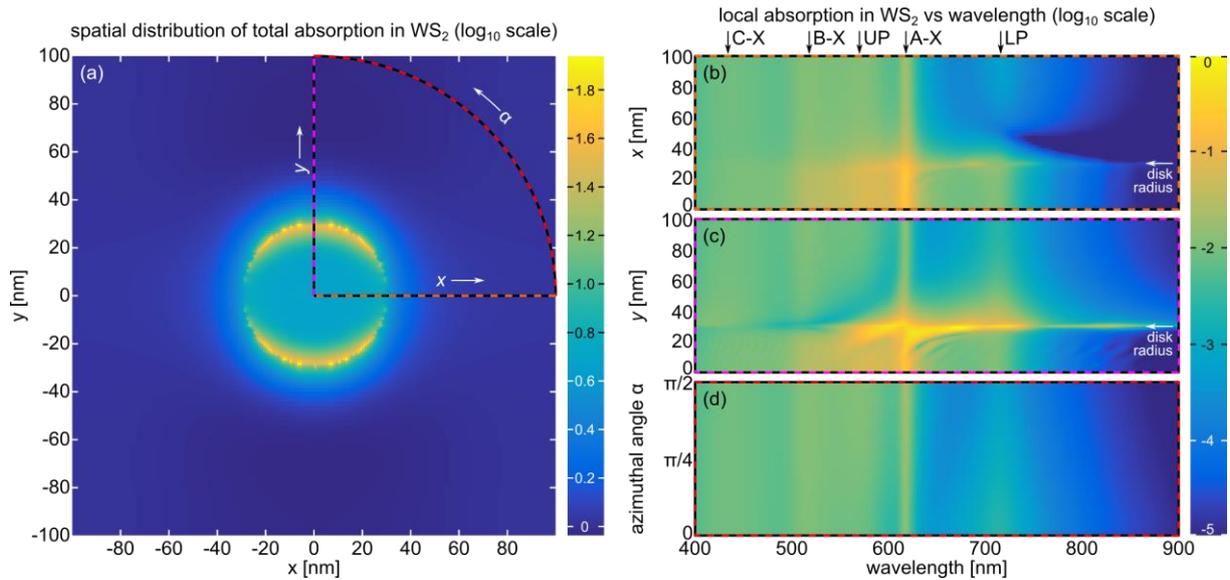

**Figure S7.** Spatially and spectrally resolved absorption in the WS$_2$ monolayer coupled with the FP cavity and nanodisk array. (a) In-plane cross section of total, wavelength-integrated absorption (assuming flat intensity spectrum; 400–900 nm range) of light in the monolayer. The influence of the localized plasmon coupled to the FP cavity is clearly visible by locally enhanced absorption near the edges of the nanodisk (using polarized light in the y-direction). The dashed lines mark spatial cross sections along which spectrally resolved absorption is plotted in (b-d). (b-d) Local spectrally resolved absorption along (b) x-cross section, (c) y-cross section, and (d) azimuthal cross section. The arrows in panel (a) mark the direction of increasing values of the vertical axes. The spectral positions of the upper and lower polaritons and excitons are marked with vertical arrows in (b). Spectral splitting of absorption is clearly visible in all cross sections.

**Supplementary References:**

1   Berkdemir, A. *et al.* Identification of individual and few layers of WS2 using Raman Spectroscopy. *Scientific Reports* **3**, 1755, doi:10.1038/srep01755 (2013).